\documentclass[conference]{IEEEtran}
\IEEEoverridecommandlockouts
\usepackage{cite}
\usepackage{amsmath,amssymb,amsfonts,amsthm}
\usepackage{algorithmic}
\usepackage{graphicx}
\usepackage{textcomp}
\usepackage{subfig}
\usepackage{xcolor}
\def\BibTeX{{\rm B\kern-.05em{\sc i\kern-.025em b}\kern-.08em
    T\kern-.1667em\lower.7ex\hbox{E}\kern-.125emX}}
\usepackage[font=small]{caption}
\usepackage[capitalize]{cleveref}
\usepackage{framed} 

\usepackage{enumerate}
\usepackage[shortlabels]{enumitem}
\newtheorem{theorem}{Theorem}
\newtheorem{corollary}{Corollary}
\newtheorem{lemma}{Lemma}
\newtheorem{proposition}{Proposition}
\theoremstyle{remark}
\newtheorem{remark}{Remark}
\begin{document}

\title{\huge Hybrid Satellite-Ground Deployments for Web3 DID: \\System Design and Performance Analysis
\thanks{The work in this paper was supported in part by the Hong Kong Metropolitan University Research Grant No. RD/2023/2.22, in part by the grant from the Research Grants Council of the Hong Kong Special Administrative Region, China, under project No. UGC/FDS16/E15/24, and in part by the Team-based Research Fund under Project No. TBRF/2024/1.10.}
\vspace{-0.2em}}

\author{
\IEEEauthorblockN{
Yalin Liu}
\IEEEauthorblockA{
\textit{Hong Kong Metropolitan University}\\
Hong Kong, China \\
ylliu@hkmu.edu.hk}
\and
\IEEEauthorblockN{
Zhigang Yan}
\IEEEauthorblockA{
\textit{Macau University of Science and Technology}\\
Macao, China \\
3220005784@student.must.edu.mo}
\and
\IEEEauthorblockN{
Bingyuan Luo}
\IEEEauthorblockA{
\textit{Hong Kong Metropolitan University}\\
Hong Kong, China \\
bentherlaw@gmail.com}
\and
\IEEEauthorblockN{
Xiaochi Xu}
\IEEEauthorblockA{
\textit{Hong Kong Metropolitan University}\\
Hong Kong, China \\
xxu@hkmu.edu.hk}
\and
\IEEEauthorblockN{
Hong-Ning Dai}
\IEEEauthorblockA{
\textit{Hong Kong Baptist University}\\
Hong Kong, China \\
hndai@ieee.org}
\and
\IEEEauthorblockN{
Yaru Fu}
\IEEEauthorblockA{
\textit{Hong Kong Metropolitan University}\\
Hong Kong, China \\
yfu@hkmu.edu.hk}
\and
\IEEEauthorblockN{
Bishenghui Tao}
\IEEEauthorblockA{
\textit{Hong Kong Metropolitan University}\\
Hong Kong, China \\
btao@hkmu.edu.hk}
\and
\IEEEauthorblockN{
Siu-Kei Au Yeung}
\IEEEauthorblockA{
\textit{Hong Kong Metropolitan University}\\
Hong Kong, China \\
jauyeung@hkmu.edu.hk}
}

\maketitle

\begin{abstract}
The emerging Web3 has great potential to provide worldwide decentralized services powered by global-range data-driven networks in the future. To ensure the security of Web3 services among diverse user entities, a decentralized identity (DID) system is essential. Especially, a user's access request to Web3 services can be treated as a DID transaction within the blockchain, executed through a consensus mechanism. However, a critical implementation issue arises in the current Web3, i.e., how to deploy network nodes to serve users on a global scale. To address this issue, emerging Low Earth Orbit (LEO) satellite communication systems, such as Starlink, offer a promising solution. With their global coverage and high reliability, these communication satellites can complement terrestrial networks as Web3 deployment infrastructures. In this case, this paper develops three hybrid satellite-ground modes to deploy the blockchain-enabled DID system for Web3 users. Three modes integrate ground nodes and satellites to provide flexible and continuous DID services for worldwide users. Meanwhile, to evaluate the effectiveness of the present hybrid deployment modes, we analyze the complete DID consensus performance of blockchain on three hybrid satellite-ground modes. Moreover, we conduct numerical and simulation experiments to verify the effectiveness of three hybrid satellite-ground modes. The impacts of various system parameters are thoroughly analyzed, providing valuable insights for implementing the worldwide Web3 DID system in real-world network environments. 

\end{abstract}



\section{Introduction}

Web3 is envisioned as the future of the Internet, providing a decentralized, secure, and open web platform~\cite{web3}. Unlike Web2, which relies on centralized entities to manage user information, Web3 leverages blockchain and decentralized networks to enable peer-to-peer interactions, ensuring transparency, security, and user autonomy. Based on decentralized networks, Web3 can provide various decentralized applications (aka Web3 services), including DeFi, DeSci, DeEco, and the Metaverse~\cite{defi}. Nowadays, benefiting from the developed blockchain technologies, Web3 services—particularly DeFi and the Metaverse—are experiencing significant growth. With data security and privacy, the current Web3 has great potential to support a wide range of user scenarios for future data-based networks, such as smart cities, agricultural Internet of Things, maritime surveillance, and geological monitoring. However, untrustworthy entities from diverse scenarios bring significant security risks in accessing Web3 services. To sum up, implementing Web3 demands a network that can provide secure Web3 services for diverse and worldwide users~\cite{web3b}.

\begin{figure}[!t]
	\centering
	\includegraphics[width=3.3in]{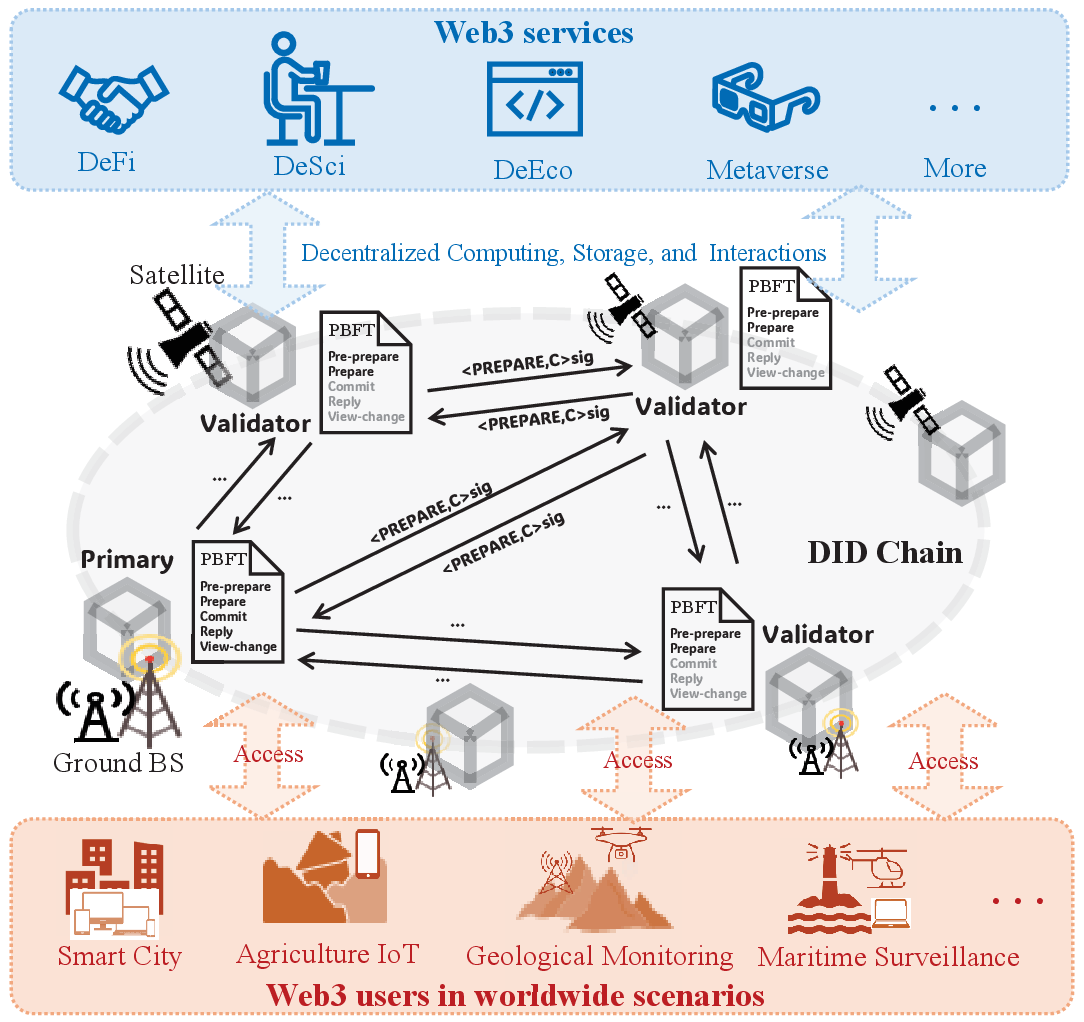}
	\caption{The blockchain-enabled Web3 architecture employs global-range network infrastructures and serves diverse user scenarios.}
        \label{fig_web3scenario}
        \vspace{-0.2cm}
\end{figure}

To provide secure Web3 services, implementing a decentralized identity (DID) system is essential~\cite{did1}. The DID system manages the identities of various Web3 entities, such as user info, wallet data, and transactions. This system significantly enhances the security and privacy of all data within Web3 services~\cite{did2,did3}. The intrinsic blockchain technology used in Web3, with its decentralized and immutable nature, offers a promising solution for managing DID. Particularly, a user's access request to Web3 services can be treated as a DID transaction within the blockchain, executed through a consensus mechanism. For simplicity, we refer to this as \textit{blockchain-enabled DID} for Web3 services. Though blockchain-enabled DID enables security in Web3, another issue remains, i.e., how to deploy network nodes to cover global-range users, especially in remote areas or situations where terrestrial infrastructures may be unavailable or compromised. To address this issue, emerging Low Earth Orbit (LEO) satellite communication systems, such as Starlink, with global-range coverage and reliability, can be deployed as complementary Web3 infrastructures to terrestrial networks~\cite{s2}. In this case, hybrid satellite-ground network infrastructures can be used to implement decentralized networks for Web3, as shown in \cref{fig_web3scenario}. Based on hybrid satellite-ground networks, \textit{the deployment scheme for blockchain-enabled DID} is required. Meanwhile, \textit{the DID performance deployed on hybrid satellite-ground chain nodes} needs to be thoroughly investigated to verify the effectiveness of the deployment scheme. 

Over the years, numerous studies have explored the deployment and performance of blockchain networks, particularly focusing on scalability, security, and the efficiency of different consensus algorithms. Consensus is the key task in blockchain to achieve a distributed agreement about the ledger's state or the state of a data set. For instance, algorithms such as Proof of Work (PoW)~\cite{pow}, Proof of Stake (PoS)~\cite{pos}, and Practical Byzantine Fault Tolerance (PBFT)~\cite{pbft1} are explored according to capability, responsibility, and availability of the chain nodes to achieve consensus. While significant progress has been made in investigating these algorithms in traditional blockchains~\cite{pa1,pa2,pa3}, few studies deploy blockchain in hybrid satellite-ground network infrastructures~\cite{s3,pa3}. It is worth noting that the blockchain performance in hybrid satellite-ground networks is hindered by factors such as network outages, limited ground infrastructure, and restricted global coverage, especially in remote or disaster-prone regions. Several solutions have been proposed to mitigate these challenges, such as using edge computing to decentralize data storage and processing~\cite{pa3} or integrating satellite communication networks with blockchain to improve the efficiency of blockchain~\cite{s3}. However, these solutions primarily focus on either improving coverage or increasing scalability in blockchain, with limited attention on the chain deployment on hybrid satellite-ground nodes for enhancing the blockchain consensus in dynamic and variable networks.

In this paper, we propose three hybrid deployment modes for Web3 DIDs, catering to dynamic and variable satellite-ground networks in practice. The three modes give the three different combinations of ground nodes and LEO satellites to provide continuous access to blockchain-enabled DID. These three modes also provide flexible choices in completing DID consensuses 
for various network situations, even during network failures or in remote areas. To serve Web3 services anytime and anywhere, the ubiquitous and reliable wireless networks among all satellite-ground chain nodes are crucial. However, the hybrid satellite-ground networks possess diverse wireless links with variable characteristics, such as stochastic propagation environments, random interference, and antenna pointing error, thus significantly affecting the final DID performance in Web3. To explore the effectiveness of three hybrid deployment modes for Web3 DID, we investigate their consensus performance under diverse wireless links. To sum up, our main contributions are summarized as follows:

\begin{itemize}
    \item We develop three hybrid satellite-ground modes to deploy the blockchain-enabled DID system for Web3 users. Three chain modes are presented by combining ground nodes and satellites to provide flexible and continuous DID services for worldwide users. The blockchain-enabled DID chain is maintained in three hybrid modes to serve secure identity management for Web3 services. 
    \item We investigate the complete DID consensus performance of blockchain on three hybrid modes. Particularly, we model and analyze the achievable value ranges of two critical metrics: the system latency and throughput, with the consideration of diverse wireless links and variable transmission environments in three modes. The analysis provides insights into the impact of comprehensive network parameters on consensus performance, thus offering useful insights for practical implementations.
    \item We conduct numerical and simulation experiments to verify the effectiveness of three hybrid satellite-ground modes and the corresponding system performance of the DID consensus. We analyze the impact of comprehensive system parameters, e.g., transmission configurations, fading and interference, and the number of chain nodes, thus giving valuable insights for implementing the DID chain in satellite-ground networks in real-world environments. 
\end{itemize}

The rest of this paper is organized as follows.~\cref{sec_systemmodel} introduces the system design and modeling.~The performance analysis of the system is given in~\cref{sec_analysis}.~\cref{sec_results} shows the numerical results.~\cref{sec_conclusion} concludes this paper.

\section{System Design and Modeling}
\label{sec_systemmodel}

\subsection{System Design} 
As illustrated in \cref{fig_web3scenario}, we propose a blockchain-enabled DID system facilitated by hybrid satellite-ground networks. The blockchain-enabled DID chain is employed to manage the identities of various Web3 entities, such as user info, wallet data, and transactions. The DID chain is deployed on the hybrid satellite-ground network, i.e., both ground and space nodes are employed to maintain the DID chain and provide Web3 services. The three hybrid modes are designed to ensure a flexible and efficient DID consensus from worldwide users. The detailed design is given as follows. 

\subsubsection{DID Chain}
\label{subsubsec_DIDchain}
In the DID chain, a user's access request to Web3 services can be treated as a DID transaction within the blockchain. Especially, a consensus mechanism in the chain is crucial in executing DID transactions and maintaining network consistency. Considering the worldwide network scale in our system, PBFT is applied since it is a relatively simple, fast, and reliable consensus mechanism. PBFT provides fast consensus by tolerating up to one-third of the network nodes being faulty or malicious~\cite{pbft1}. Under PBFT, the chain node that receives the DID request from the user is called the primary node and all of the other chain nodes are validators. A PBFT-enabled DID consensus includes the four following phases: 

    {
    \small
\begin{itemize}[leftmargin = 0em]
    \item[] \textbf{(Phase 1) Pre-Prepare:} The primary node sends a \texttt{PRE-PREPARE} message containing a DID transaction to the validators (i.e. other chain nodes).
    \item[] \textbf{(Phase 2) Prepare:} Each validator verifies the transaction and sends a \texttt{PREPARE} message to other validators if it agrees with the transaction.
    \item[] \textbf{(Phase 3) Commit:} Once enough \texttt{PREPARE} messages are received, i.e., more than $2/3$ of the validators, the validator sends a \texttt{COMMIT} message. 
    \item[] \textbf{(Phase 4) Reply:} After receiving sufficient \texttt{COMMIT} messages, the transaction is considered finalized and the DID consensus is executed. All validators send a \texttt{REPLY} message to the user, confirming the DID request's completion.
\end{itemize}
}

\begin{figure}[!t]
	\centering
	\includegraphics[width=3.6in]{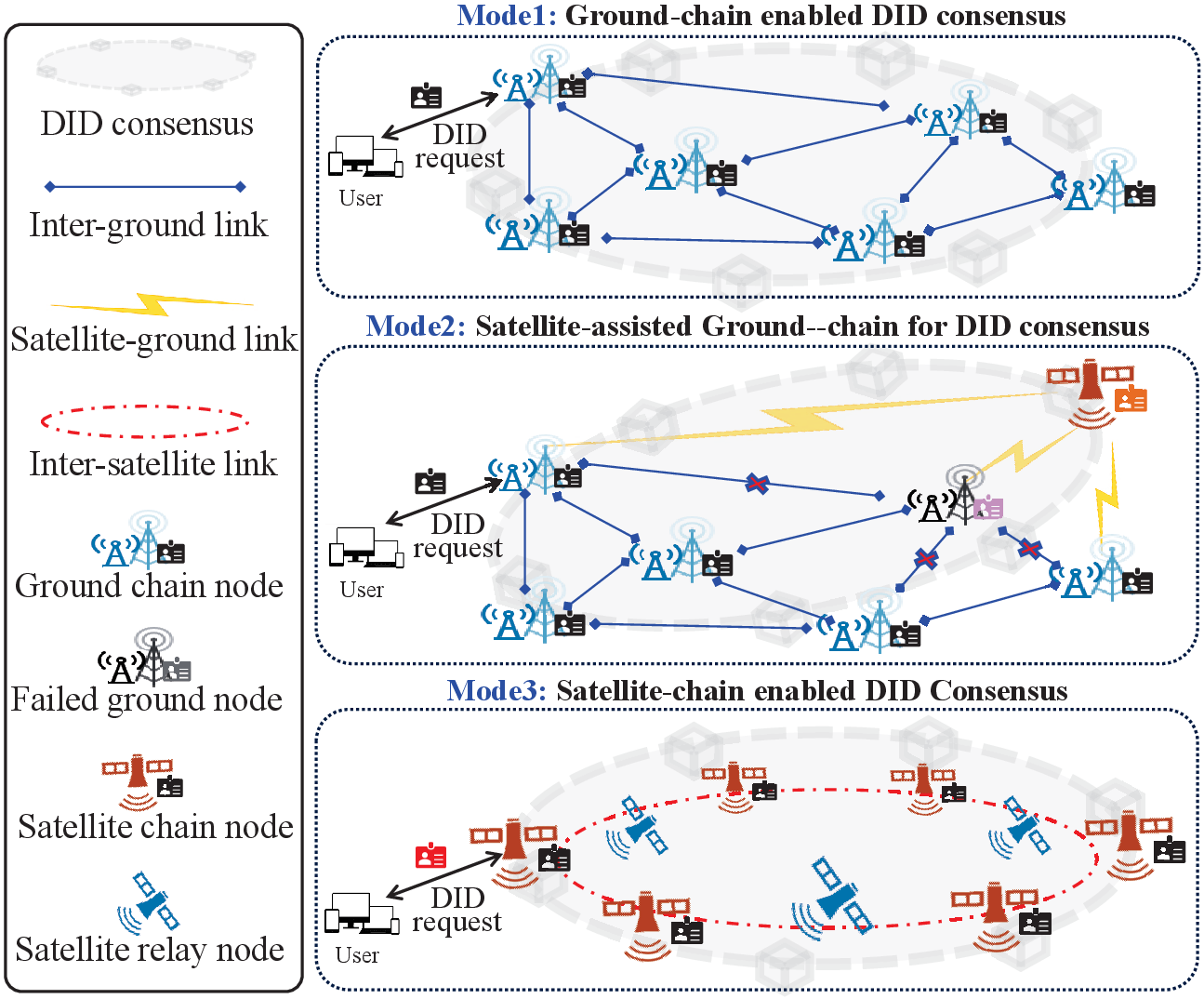}
	\caption{DID consensuses under three hybrid satellite-ground modes.}
        \label{fig_hybridmode}
\end{figure}

\subsubsection{Satellite-ground Nodes}
Considering the intrinsic characteristics of ground and satellite nodes, they work in the DID system in different roles. \textit{First}, the ground-side chain nodes (called ground chain nodes) are deployed on trustful ground infrastructures, e.g., operated by governments and enterprises, capable of hosting the primary blockchain functionalities and executing consensus mechanisms to complete DID validation tasks. \textit{Second}, the space-side chain nodes (called satellite chain nodes) are LEO communication satellite constellations with massive access capabilities, e.g., Starlink, serving as an assisted blockchain layer, providing backup verification capabilities when ground nodes are unavailable. \textit{Besides}, some normal satellites (called satellite relay nodes) are also used to support relay transactions, store temporary identity proofs, and initiate consensus when required, ensuring the global availability of identity services and acting as failover mechanisms.

\subsubsection{Hybrid Modes}
\label{subsubsec_modes}
Due to the availability and mobility of network nodes, users may submit their access requests to different chain nodes, i.e., a ground BS or a satellite. In this case, we design a flexible and efficient DID consensus through three modes of hybrid satellite-ground chains. As shown in~\cref{fig_hybridmode}, three hybrid modes adapt different access points and chain nodes to ensure users request DID services anywhere, even in remote areas or during network outages. We give the detailed design of the three modes as follows:

{\small
\begin{itemize}[leftmargin=*]
    \item \textbf{Mode 1:} \textit{Ground-chain enabled DID consensus.} DID requests are processed entirely on the ground node-enabled blockchain (called Ground-chain). This mode works when the available ground nodes are enough to complete consensus. Thus, satellites are not required. In this mode, satellite nodes may need to synchronize some important transaction data periodically for backup requirements. 
    \item \textbf{Mode 2:} \textit{Satellite-assisted Ground-chain for DID consensus.} DID requests are processed on a Ground-chain that is assisted by satellites. This mode works when partial ground nodes are temporarily offline due to network failures, resulting in insufficient chain nodes to complete consensus. Then satellites join as assistant chain nodes to retrieve DID transactions and provide validation services. 
    \item \textbf{Mode 3:} \textit{Satellite-chain enabled DID consensus.} 
    DID requests are processed entirely on the satellite node enabled blockchain (called Satellite-chain). This mode is used when ground nodes are completely disconnected due to disasters, attacks, or geographical isolation. LEO satellites can form a satellite chain that processes DID validation via the blockchain consensus. During idle periods, the satellite-chain need to send all transaction data to the restored or other Ground-chains for data relief and records.
\end{itemize}
}
Among all three hybrid modes of the DID chain,
there are three types of transmission links: i) \textit{inter-ground links} in \textbf{Mode 1} and \textbf{Mode 2}; ii) \textit{satellite-ground links} in \textbf{Mode 2}; and iii) \textit{inter-satellite links} in \textbf{Mode 3}. The transmission performance of these links directly affects the overall DID chain performance, which will be given in~\cref{sec_analysis}.

\subsection{DID Performance Modeling}
\label{subsec_DIDperformance}
We evaluate the overall performance of the DID chain in two typical system metrics, i.e., throughput and latency~\cite{did3}. Considering a DID chain that includes the node set $\mathcal{X}$, let $x_{\mathrm{0}}\in\mathcal{X}$ and $(x_{\mathrm{t}},x_{\mathrm{r}})\in\mathcal{X}$ be the primary node and any transceiver pair of a point-to-point communication in the PBFT consensus, respectively. Let $\mathsf{TPS}$ and $T_{\mathrm{overall}}$ be the system throughput and latency of the DID chain, respectively. $\mathsf{TPS}$ is determined by the number of transactions completed per second. $T_{\mathrm{overall}}$ is the total latency of completing a transaction, which is mainly determined by the consensus latency of the PBFT process. 
The values of $\mathsf{TPS}$ and $T_{\mathrm{overall}}$ are given in~\cref{lem1}.
\begin{lemma}
    Given the node set $\mathcal{X}$, the primary node $x_{\mathrm{0}}\in\mathcal{X}$ and any transceiver pair $(x_{\mathrm{t}},x_{\mathrm{r}})\in\mathcal{X}$ of two nodes in the DID chain, the values of $\mathsf{TPS}$ and $T_{\mathrm{overall}}$ are given by  
    \vspace{-0.3cm}

    {\footnotesize
    \begin{align*}
        &\mathsf{TPS}=\frac{1}{T_{\mathrm{overall}}},T_{\mathrm{overall}}=\sum_{i=1}^4T_i,\\
        &T_i=\left\{ 
            \begin{matrix}
             \max\{ t_c(x_{\mathrm{t}},x_{\mathrm{r}})\},&\text{if } i=\{1,2,3\}.\\
             \max\{ t_c(x_{\mathrm{t}},x_{\mathrm{0}})\},&\text{if } i=4.\\
            \end{matrix}
        \right. 
    \end{align*}}%
    The terms $T_i (i\in\{1,2,3,4\})$ and $t_c(\cdot)$ are is the latency of the $i$-th phase during the PBFT consensus process and the communication latency of two nodes in $\mathcal{X}$, respectively. 
    \label{lem1}
\end{lemma}
\vspace{-0.2cm}
\noindent \textit{Proof.} The proof process is given in Appendix~\ref{proof_lem1}. \hfill $\blacksquare$

Considering wireless communication in practice, the latency $t_c$ among two nodes of the DID chain is affected by stochastic channel fading and interference. The stochastic wireless link performance among two nodes is usually modeled by the successful transmission probability, denoted by $P_{s}$. According to the relationship between $t_c$ and $P_{s}$, we have~\cref{lem2}.

\begin{lemma}
    For a specific wireless link between two nodes in the DID chain, $P_{s}$ is the successful transmission probability, $N$ is the number of orthogonal subcarriers, $B$ is the bandwidth of each subcarrier, $C$ is the Shannon capacity, and $R$ is the available transmission rate. The value of $t_c$ is given by
    \vspace{-0.3cm}

{\footnotesize
    \begin{align*}
        &t_c=\Phi^{-1}\left(Q^{-1}(1{-}P_{s})\right),\\
        &\Phi(x)=\frac{(C-R)BNx+0.5\log_{2}BNx}{\log_{2}e\cdot \sqrt{BNx}},
    \end{align*}}%
    where $Q(\cdot)$ is the $Q$-function.
    \label{lem2}
\end{lemma}%
\vspace{-0.2cm}
\noindent \textit{Proof.} The proof process is given in Appendix~\ref{proof_lem2}.
\hfill $\blacksquare$

\section{Theoretical Analysis}
\label{sec_analysis}
In this section, we analyze the overall performance of the proposed DID chain, i.e., $\mathsf{TPS}$ and $T_{\mathrm{overall}}$. Referring to~\cref{lem1}, $\mathsf{TPS}$ is directly determined by $T_{\mathrm{overall}}$, so we focus on the theoretical analysis of $T_{\mathrm{overall}}$. Since our DID chain is deployed in three hybrid satellite-ground modes. We need to investigate $T_{\mathrm{overall}}$ for the three modes, respectively. As given in~\cref{lem1,lem2}, $T_{\mathrm{overall}}$ of three modes can be calculated by substituting $P_{s}$ and $t_c$ of transmission links in the corresponding mode. Referring to~\cref{lem2}, $t_c$ of a transmission link can be calculated by $P_{s}$. Therefore, we first model $P_{s}$ for different transmission links in~\cref{subsec_ps}. Substituting $P_{s}$ into~\cref{lem1,lem2}, we can analyze $T_{\mathrm{overall}}$ and $\mathsf{TPS}$ of three modes in~\cref{subsec_ps}.
\vspace{-0.2em}

\subsection{$P_{s}$ Analysis of Transmission Links}
\label{subsec_ps}
Our hybrid deployment modes include three types of transmission links, i.e., \textit{inter-ground links}, \textit{satellite-ground links}, and \textit{inter-satellite links}. Let $l=\{\mathrm{g},\mathrm{sg},\mathrm{s}\}$ be the index of the three link types, respectively. The general form of $P_{s,l},t_{c,l}$ for three link types can be given in~\cref{pro0}.

\begin{proposition}
    For the link type $l$, $\forall l=\{\mathrm{g},\mathrm{sg},\mathrm{s}\}$, $\gamma_{l}$ is the received Signal-to-noise ratio (SNR) or Signal-to-interference-plus-noise ratio (SINR), and $\gamma_{0,l}$ is the detection threshold of SNR or SINR at the receiver. The successful transmission probability $P_{s,l}$ and the link latency $t_{c,l}$ are given by
    \vspace{-0.3cm}

{\footnotesize
    \begin{align*}
        &P_{s,l}=
            \begin{cases}
            \mathbb{P}(\gamma_{l}\geq \gamma_{0,l}),\quad\quad \textrm{Interference-free case}.\\
            \int_{0}^{D_{l}}\mathbb{P}(\gamma_{l}\geq \gamma_{0,l})f(d)\mathrm{d}d, \textrm{Interference case}.
            \end{cases}\\
        &t_{c,l}=\Phi^{-1}\left(Q^{-1}(1{-}P_{s,l})\right),
    \end{align*}}%
    where $d$ is the distance from the interfering node to the receiver, $D_{l}$ is the maximum detection distance of the interference, 
    $Q(\cdot)$ is the $Q$-function, and $\Phi(\cdot)$ is given in~\cref{lem2}.
    \label{pro0}
\end{proposition}
\vspace{-0.6cm}
\noindent \textit{Proof.} The proof process is given in Appendix~\ref{proof_pro1}.
\hfill $\blacksquare$
\begin{remark}
    In~\cref{pro0}, $t_{c,l}$ is determined by $P_{s,l}$. 
    Due to the diverse transmission characteristics in different links, obtaining closed-form expressions of $P_{s,l}$ for all links is challenging. To address this issue, we evaluate the upper and lower bounds of $P_{s,l}$ for some transmission links as follows.
    \label{remark_pro0}
\end{remark}

\subsubsection{$P_{s,\mathrm{g}}$ for Inter-Ground Links}
\label{subsubsec_mode1analysis}
Inter-ground links are used to build transmissions among ground chain nodes in both \textbf{Mode 1} and \textbf{Mode 2} of our hybrid DID chain. Due to the spectrum constraint and dense communications in terrestrial networks, the inter-ground links are greatly affected by the interference. Given a typical inter-ground link with the transmission distance $d_0$, the SINR is given by~\cite{pa2} 
\vspace{-0.3cm}

{\footnotesize
\begin{align}
    &\gamma_{\mathrm{g}}=\frac{p_{\mathrm{g}}L_{\mathrm{g}}(d_0)}{\sum_{\Phi_{\mathrm{g}}(D_{\mathrm{g}})}p_{\mathrm{g}}L_{\mathrm{g}}(d_x)+\delta_{\mathrm{g}}^2},\label{eq_gamma_g}
\end{align}}%
where $d_{0}$ is the transmission distance between two nodes, $p_{\mathrm{g}}$ is the fixed transmission power for all ground nodes, $L_{\mathrm{g}}(d)=(c/4\pi d f_{\mathrm{g}})^2$ is the path loss function for inter-ground links with $c$ being the speed of the light and $f_{\mathrm{g}}$ being the used carrier frequency, $\Phi_{\mathrm{g}}(D_{\mathrm{g}})$ is the interferer node-set,  and $\delta_{\mathrm{g}}$ is the additive noise power. We consider that ground nodes in the terrestrial network are distributed as a Homogenous Poisson Point Process (HPPP) with the intensity $\lambda_{\mathrm{g}}$. Let $S_{\mathrm{g}}=p_{\mathrm{g}}L_{\mathrm{g}}(d_0)$, $I_{\mathrm{g}}=\sum_{\Phi_{\mathrm{g}}(D_{\mathrm{g}})}p_{\mathrm{g}}L_{\mathrm{g}}(d_x)$ be the desired signal power and the interference, respectively. The probability density functions (PDF) of $\gamma_{\mathrm{g}}$, $S_{\mathrm{g}}$, and $I_{\mathrm{g}}$ are given by~\cite{pa2}.
\vspace{-0.3cm}

{\footnotesize
\begin{align}
    &f_{\gamma_{\mathrm{g}}}=f_{S_{\mathrm{g}}}f_{I_{\mathrm{g}}},
    f_{S_{\mathrm{g}}}=2\pi\lambda_{\mathrm{g}}d_0e^{-\lambda_{\mathrm{g}}\pi d_0^{2}},
    f_{I_{\mathrm{g}}}=\left(\frac{2}{D_{\mathrm{g}}^{2}}\right)^{K}\prod_{\Phi_{\mathrm{g}}(D_{\mathrm{g}})}d_x\label{eq_fi}
    \end{align}}%
where $K=|\Phi_{\mathrm{g}}(D_{\mathrm{g}})|=\left \lceil \pi \lambda_{\mathrm{g}} D_{\mathrm{g}}^2\right \rceil $ is the mean number of interferer nodes in the set $\Phi_{\mathrm{g}}(D_{\mathrm{g}})$. Substituting~\cref{eq_gamma_g,eq_fi} into~\cref{pro0}, we can derive the upper and lower bound of $P_{s,\mathrm{g}}$, as given in~\cref{cro1}.

\begin{corollary}
    Let $P_{s,\mathrm{g}}^{\mathrm{L}}$ and $P_{s,\mathrm{g}}^{\mathrm{U}}$ be the \underline{L}ower bound and \underline{U}pper bound of $P_{s,\mathrm{g}}$ for inter-ground links, which are given by
    \vspace{-0.3cm}

    {\footnotesize
    \begin{align*}
        &P_{s,\mathrm{g}}^{\mathrm{L}}=\left(1-\frac{D_{\mathrm{g}}^{2}}{4}\left(\frac{e^{\pi\lambda_{\mathrm{g}} d_{u}^{2}}}{(\gamma_{0,\mathrm{g}}-\varepsilon)2\pi\lambda_{\mathrm{g}}d_u}\right)^{\frac{2}{\left \lceil \pi \lambda_{\mathrm{g}} D_{\mathrm{g}}^2\right \rceil}}\right)^{\left \lceil \pi \lambda_{\mathrm{g}} D_{\mathrm{g}}^2\right \rceil},\\
        &P_{s,\mathrm{g}}^{\mathrm{U}}=1-\Bigg(\frac{D_{\mathrm{g}}^{2}}{4}\Bigg)^{\left \lceil \pi \lambda_{\mathrm{g}} D_{\mathrm{g}}^2\right \rceil}\left(\frac{(1-\varepsilon)e^{\pi\lambda_{\mathrm{g}} d_{l}^{2}}}{(\gamma_{0,\mathrm{g}}-\varepsilon)2\pi\lambda_{\mathrm{g}}d_l}\right)^{2},
    \end{align*}
    }
    where $\varepsilon\leq\gamma_{0,\mathrm{g}}$ is a positive constant, $d_{l}$ and $d_{u}$ are the minimal and maximal distance among all ground nodes.
    \label{cro1}
\end{corollary}\vspace{-0.2cm}
\noindent \textit{Proof.} The proof process is given in Appendix~\ref{proof_cro1}.\hfill$\blacksquare$

\begin{remark}
    As given in~\cref{cro1}, $P_{s,\mathrm{g}}^{\mathrm{L}}$ and $P_{s,\mathrm{g}}^{\mathrm{U}}$ are affected by the interference distribution parameters 
    $\{\lambda_{\mathrm{g}},D_{\mathrm{g}}\}$, the transmission parameters $\{d_0,\gamma_{0,\mathrm{g}}\}$, and the bound factor $\varepsilon$. 
    The interference distribution affects a lot on $P_{s,\mathrm{g}}^{\mathrm{L}}$ and $P_{s,\mathrm{g}}^{\mathrm{U}}$, making it hard to see the monotonic relationship of system parameters. Even though, we can observe that $P_{s,\mathrm{g}}^{\mathrm{L}}$ and $P_{s,\mathrm{g}}^{\mathrm{U}}$ increase with the increasing $\gamma_{0,\mathrm{g}}$ and the decreasing $\varepsilon$. The increasing of $\gamma_{0,\mathrm{g}}$ improves the detection threshold of the received signal power, thus reducing 
    interference. Recall that $\varepsilon$ is a constant smaller than $\gamma_{0,\mathrm{g}}$, thus can be used to adjust the bound sensitivity.
    \label{remark_cro1}
\end{remark}

\subsubsection{$P_{s,\mathrm{sg}}$ for Satellite-Ground Links}
\label{subsubsec_mode2analysis}
In our hybrid DID chain, satellite-ground links are used in \textbf{Mode 2} to synchronize data from the ground chain to satellites. For efficient synchronization, we consider that one ground node from the ground chain is chosen, as a representative, to build the transmission with the satellite. Given a typical satellite-ground link with the transmission distance $d_0$, the SNR is given by 
\vspace{-0.3cm}

{\footnotesize
\begin{align}
    \gamma_{\mathrm{sg}}=\frac{p_{\mathrm{sg}}G_{\mathrm{sg}}|h_{\mathrm{sg}}|^{2}L_{\mathrm{sg}}(d_{\mathrm{sg}})}{(\sigma_{\mathrm{sg}})^{2}},\label{eq_gamma_sg}
\end{align}}%
where $p_{\mathrm{sg}}$ is the fixed transmission power, $G_{\mathrm{sg}}$ is the effective antenna gain, $L_{\mathrm{sg}}(d)=(c/4\pi d f_{\mathrm{sg}})^2$ is the path loss function for satellite-ground links with $f_{\mathrm{sg}}$ being the used carrier frequency, and $\delta_{\mathrm{sg}}$ is the additive noise power. The term $|h_{\mathrm{sg}}|^2$ is the channel fading gain of the satellite-ground link. According to the Shadowed-Rician fading model for the satellite-ground channel fading~\cite{sg}, the PDF of $|h_{\mathrm{sg}}|^2$ is given by
\vspace{-0.2cm}

{\footnotesize
\begin{equation}
      f_{|h_{\mathrm{sg}}|^2}(x)=\frac{1}{\beta^{\alpha}\Gamma(\alpha)}x^{\alpha-1}e^{-\frac{x}{\beta}},\label{eq_fh}
\end{equation}}%
where $\Gamma(\cdot)$ is the gamma function and $\alpha,\beta$ are the shape and scale parameters of the channel fading variable $h_{\mathrm{sg}}$. \footnote{The values of $\alpha=\frac{m_{0}(2b_{0}+\Omega)^{2}}{4m_{0}b_{0}^{2}+4m_{0}b_{0}\Omega+\Omega^{2}}$ and $\beta=\frac{2b_{0}+\Omega}{\alpha}$, which are determined by the Nakagami fading coefficient, the average power of scattered components, and the average power of the line-of-sight component~\cite{rice}.}. 
Substituting~\cref{eq_gamma_sg,eq_fh} into~\cref{pro0}, we have~\cref{cro2}.
\begin{corollary}
$P_{s,\mathrm{sg}}$ for satellite-ground links is given by
\vspace{-0.3cm}

{\small
\begin{align*}
    &P_{s,\mathrm{sg}}
    =1-\frac{\varrho\left(\alpha,\frac{\gamma_{0,\mathrm{sg}}(\sigma_{\mathrm{sg}})^{2}}{\beta p_{\mathrm{sg}}G_{\mathrm{sg}}L_{\mathrm{sg}}(d_{\mathrm{sg}})}\right)}{\Gamma(\alpha)},
    L_{\mathrm{sg}}(d_0)=\left(\frac{c}{4\pi d_{\mathrm{sg}}f_{\mathrm{sg}}}\right)^2,
\end{align*}}%
where $\varrho(\cdot,\cdot)$ is the lower incomplete gamma function, $\Gamma(\cdot)$ is the gamma function, and $d_{\mathrm{sg}}$ is the distance between the satellite and the ground node. 
    \label{cro2}
\end{corollary}\vspace{-0.2cm}
\noindent \textit{Proof.} The proof process is given in Appendix~\ref{proof_cro2}.\hfill$\blacksquare$

\begin{remark}
    As given in~\cref{cro2}, $P_{s,\mathrm{sg}}$ is affected by the channel fading parameters $\{\alpha,\beta\}$ and the transmission parameters $\{p_{\mathrm{sg}},G_{\mathrm{sg}},\gamma_{0,\mathrm{sg}},\sigma_{\mathrm{sg}},d_0,f_{\mathrm{sg}}\}$. Since the lower incomplete gamma function $\varrho(\cdot,x)$ is a increasing function of $x$, $P_{s,\mathrm{sg}}$ is a decreasing function of $\gamma_{0,\mathrm{sg}}(\sigma_{\mathrm{sg}})^{2}/(\beta p_{\mathrm{sg}}G_{\mathrm{sg}}L_{\mathrm{sg}}(d_0))$. Thereby, we observed that, $P_{s,\mathrm{sg}}$ increases with the increasing of $\{\beta,p_{\mathrm{sg}},G_{\mathrm{sg}}\}$ and the decreasing of $\{\gamma_{0,\mathrm{sg}},\sigma_{\mathrm{sg}},d_0,f_{\mathrm{sg}}\}$.
    \label{remark_cro2}
\end{remark}

\subsubsection{$P_{s}$ for Inter-Satellite Links}
\label{subsubsec_mode3analysis}
Inter-satellite links are used to build line-of-sight transmissions of satellite chain nodes in \textbf{Mode 3} of our hybrid chain. Given a typical inter-satellite link with the transmission distance $d_0$, the SNR is given by 
\vspace{-0.3cm}

{\footnotesize
\begin{align}
    \gamma_{\mathrm{s}}=\frac{p_{\mathrm{s}}G_{\mathrm{s}}W_{\mathrm{s}}L_{\mathrm{s}}(d_0)}{(\sigma_{\mathrm{s}})^{2}},\label{eq_gamma_s}
\end{align}}%
where $p_{\mathrm{s}}$ is the fixed transmission power, $G_{\mathrm{s}}$ is the effective antenna gain, $W_{\mathrm{s}}$ is path loss caused by the antenna pointing error, $L_{\mathrm{s}}(d)=(c/4\pi d f_{\mathrm{s}})^2$ is the path loss function for inter-satellite links with $f_{\mathrm{s}}$ being the used carrier frequency, and $\delta_{\mathrm{s}}$ is the additive noise power.
In practice, the beam pointing or two satellites has deviation, leading to the pointing error. Let $\theta_{d}$ be the beam deviation angle 
that follows a Rayleigh distribution with variance $\varsigma$. 
The PDF of $\theta_{d},W_{\mathrm{s}}$ is given by~\cite{point}
\vspace{-0.3cm}

{\footnotesize
\begin{align}
    &f_{\theta_{d}}(\theta_{d})=\frac{\theta_{d}}{\varsigma^{2}}e^{-\frac{\theta_{d}}{2\varsigma^{2}}},
    f_{W_{\mathrm{s}}}(w)=\frac{\eta_{s}^{2}w^{\eta_{s}^{2}-1}\cos{\theta_{d}}}{A_{0}^{\eta_{s}^{2}}},w\in[0,A_{0}] \label{eq_fw}
\end{align}}%
where $\eta_{s}$ and $A_{0}$ are antenna pointing parameters~\cite{point}. Substituting~\cref{eq_gamma_s,eq_fw} into~\cref{pro0}, we have~\cref{cro3}.
\begin{corollary}
    Let $P_{s,\mathrm{s}}^{\mathrm{L}}$ and $P_{s,\mathrm{s}}^{\mathrm{U}}$ be the \underline{L}ower bound and \underline{U}pper bound of $P_{s,\mathrm{s}}$ for inter-satellite links, which are given by
    \vspace{-0.3cm}

    {\small
    \begin{align*}
        &P_{s,\mathrm{s}}^{\mathrm{L}}=(1-\varsigma^{2})\left(1-\left(\frac{\gamma_{0,\mathrm{s}}\sigma_{\mathrm{s}}^{2}}{A_{0}p_{\mathrm{s}}G_{\mathrm{s}}L_{\mathrm{s}}(d_{0}^{\max})}\right)^{\eta_{s}^{2}}\right),\\
        &P_{s,\mathrm{s}}^{\mathrm{U}}=1-\varsigma^{2},L_{\mathrm{s}}(d_{0}^{\max})=\left(\frac{c}{4\pi d_{0}^{\max} f_{\mathrm{s}}}\right)^2,\\
        &d_{0}^{\max}=\sqrt{h_{i}(h_{i}+2R_{\mathrm{e}})}+\sqrt{h_{j}(h_{j}+2R_{\mathrm{e}})},
    \end{align*}}%
    where $R_{\mathrm{e}}$ is the earth radius, and $h_{i},h_{j}$ are the altitudes of two satellites $(i,j)$ in the inter-satellite link.
  \label{cro3}
\end{corollary}%
\noindent \textit{Proof.} The proof process is given in Appendix~\ref{proof_cro3}.\hfill$\blacksquare$

\begin{remark}
    As given in~\cref{cro3}, $P_{s,\mathrm{s}}^{\mathrm{L}}$ and $P_{s,\mathrm{s}}^{\mathrm{U}}$ are affected by the pointing error parameters $\{\varsigma,\eta_{s},A_0\}$ and the transmission parameters $\{p_{\mathrm{s}},G_{\mathrm{s}},\gamma_{0,\mathrm{s}},\sigma_{\mathrm{s}},d_{0}^{\max}, f_{\mathrm{s}}\}$. Particularly, we observe that $P_{s,\mathrm{s}}^{\mathrm{L}}$ and $P_{s,\mathrm{s}}^{\mathrm{U}}$ increase with the increasing $\{A_0,p_{\mathrm{s}},G_{\mathrm{s}}\}$ and the decreasing $\{\varsigma,\eta_{s},\gamma_{0,\mathrm{s}},\sigma_{\mathrm{s}},d_{0}^{\max}, f_{\mathrm{s}}\}$. 
\label{remark_cro3}
\end{remark}

\subsection{$T_{\mathrm{overall}}$ and $\mathsf{TPS}$ Analysis of Three Modes}
\label{subsec_latencyanalysis}
In our chain system, the DID consensus in different hybrid modes is completed via different transmission links, thus leading to different latency $T_{\mathrm{overall}}$ and throughput $\mathsf{TPS}$. However, it is hard to obtain the closed-form expressions of $T_{\mathrm{overall}}$ and $\mathsf{TPS}$ directly. Thus, we derive their upper and lower bound and use this gap to estimate $T_{\mathrm{overall}}$ and $\mathsf{TPS}$.

Let $k=\{1,2,3\}$ represents \textbf{Mode 1}, \textbf{Mode 2}, and \textbf{Mode 3}, respectively. Substituting the value or value range of $P_{s}$ in corresponding links (given in~\cref{subsec_ps}) into~\cref{lem1,lem2} of three modes, the value range of $T_{k,\mathrm{overall}}$ $\mathsf{TPS}_{k}$ for three modes $k=\{1,2,3\}$ can be evaluated by~\cref{the1}.

\begin{theorem}
    Let $T_{k,\mathrm{overall}}^{z},\mathsf{TPS}_{k}^{z}, k=\{1,2,3\}, z=\{\mathrm{L},\mathrm{U}\}$ denote the \underline{L}ower bound and \underline{U}pper bound of the system latency and the system throughput of completing DID consensus in three modes. The values of $T_{k,\mathrm{overall}}^{z},\mathsf{TPS}_{k}^{z}$ are given by
    \vspace{-0.3cm}

    {\footnotesize
    \begin{align*}
    &\mathsf{TPS}_{k}^{z}=
    \left\{  
    \begin{matrix}
      \frac{1}{T_{k,\mathrm{overall}}^{\mathrm{U}}}, & \text{if } z=\mathrm{L}.\\
      \frac{1}{T_{k,\mathrm{overall}}^{\mathrm{L}}}, & \text{if } z=\mathrm{U}.
    \end{matrix}
    \right., T_{k,\mathrm{overall}}^{z}=\\
    &
    \left\{  
    \begin{matrix}
      4\Phi^{-1}\left(Q^{-1}(1{-}P_{s,\mathrm{g}}^{z})\right), & \text{if } k=1.\\
      4\max\{\Phi^{-1}\left(Q^{-1}(1{-}P_{s,\mathrm{g}}^{z})\right),\Phi^{-1}\left(Q^{-1}(1{-}P_{s,\mathrm{sg}})\right)\}, & \text{if } k=2.\\
      4\Phi^{-1}\left(Q^{-1}(1{-}P_{s,\mathrm{s}}^{z})\right), & \text{if } k=3.
    \end{matrix}
    \right.
    \end{align*}
    }%
where $P_{s,\mathrm{g}}^{z},P_{s,\mathrm{sg}},P_{s,\mathrm{s}}^{z}$ are given in~\cref{cro1,cro2,cro3}.
\label{the1}
\end{theorem}

\noindent \textit{Proof.} The proof process is given in Appendix~\ref{proof_the1}.\hfill$\blacksquare$

\begin{remark}
As given in~\cref{the1} $\Phi^{-1}(x)$ and $Q^{-1}(x)$ are increasing and decreasing functions of $x$ respectively. Thus, the increasing of $P_{s}^{U}$ in corresponding links of the mode $k$ leads to the decreasing $T_{k,\mathrm{overall}}^{L}$ and increasing $\mathsf{TPS}_{k}^{U}$. Similarly, a smaller $P_{s}^{L}$ means a larger $T_{k,\mathrm{overall}}^{U}$ and decreasing $\mathsf{TPS}_{k}^{L}$. 
Meanwhile, the gap between the upper and lower bounds is affected by comprehensive system parameters. Combining
~\cref{cro1,cro2,cro3,the1}, we summarize the impacts of different parameters to $\{T_{k,\mathrm{overall}}^{z},\mathsf{TPS}_{k}^{z}\}$. 
\begin{itemize}[leftmargin=*]
\item \textit{Impacts of the parameters on \textit{inter-ground links}:} The interference distribution parameters 
    $\{\lambda_{\mathrm{g}},D_{\mathrm{g}}\}$, the transmission parameters $\{d_0,\gamma_{0,\mathrm{g}}\}$, and the bound factor $\varepsilon$ in $P_{s,\mathrm{g}}^{z}$ affect $\{T_{1,\mathrm{overall}}^{z},\mathsf{TPS}_{1}^{z},T_{2,\mathrm{overall}}^{z},\mathsf{TPS}_{2}^{z}\}$. Referring to~\cref{remark_cro1}, the increasing of $\gamma_{0,\mathrm{g}}$ and $\varepsilon$ can lead to the decreasing $\{T_{1,\mathrm{overall}}^{z},T_{2,\mathrm{overall}}^{z}\}$ and the increasing $\{\mathsf{TPS}_{1}^{z},\mathsf{TPS}_{2}^{z}\}$.
\item \textit{Impacts of the parameters on \textit{satellite-ground links}:} The channel fading parameters $\{\alpha,\beta\}$ and the transmission parameters $\{p_{\mathrm{sg}},G_{\mathrm{sg}},\gamma_{0,\mathrm{sg}},\sigma_{\mathrm{sg}},d_0,f_{\mathrm{sg}}\}$ in $P_{s,\mathrm{sg}}$ affect $\{T_{2,\mathrm{overall}}^{z},\mathsf{TPS}_{2}^{z}\}$. Referring to~\cref{remark_cro2}, the increasing of $\{\beta,p_{\mathrm{sg}},G_{\mathrm{sg}}\}$ or the increasing of $\{\gamma_{0,\mathrm{sg}},\sigma_{\mathrm{sg}},d_0,f_{\mathrm{sg}}\}$ can lead to the decreasing $T_{2,\mathrm{overall}}^{z}$ and the increasing $\mathsf{TPS}_{2}^{z}$. 
\item \textit{Impacts of the parameters on \textit{inter-satellite links}:} The pointing error parameters $\{\varsigma,\eta_{s},A_0\}$ and the transmission parameters $\{p_{\mathrm{s}},G_{\mathrm{s}},\gamma_{0,\mathrm{s}},\sigma_{\mathrm{s}},d_{0}^{\max}, f_{\mathrm{s}}\}$ in $P_{s,\mathrm{s}}$ affect $\{T_{3,\mathrm{overall}}^{z},\mathsf{TPS}_{3}^{z}\}$. Referring to~\cref{remark_cro3}, the increasing $\{A_0,p_{\mathrm{s}},G_{\mathrm{s}}\}$ or the decreasing $\{\varsigma,\eta_{s},\gamma_{0,\mathrm{s}},\sigma_{\mathrm{s}},d_{0}^{\max}, f_{\mathrm{s}}\}$ can lead to the decreasing $T_{3,\mathrm{overall}}^{z}$ and the increasing $\mathsf{TPS}_{3}^{z}$.
\end{itemize}
\label{remark_the1}
Besides the impacts from the above parameters, 
some other parameters, such as the number of chain nodes, cannot be observed by this theorem directly, how these values affect the system latency is explored in our simulation results. Moreover, it is worth noting that, the changing trend of the upper and lower bounds cannot directly reflect the changing trend of the latency and throughput themselves. The gap between the upper and lower bounds reflects the error range of using them to estimate the latency and throughput. Our simulation result can test the impact of system parameters on latency and verify the effectiveness of the bound we proposed.
\end{remark}

\begin{table}[!t]
	\caption{Settings of Simulation Parameters\label{tab:table1}}
	\centering
	\begin{tabular}{c|c||c|c}
		\hline
		Parameter & Value & Parameter  & Value\\
		\hline
		$\delta_{g}^{2}$ & $-104$dBm~\cite{pa2} & $D_{\mathrm{g}}$ & $100$m\\
        $d_{u}$ & $80$m & $d_{l}$ & $20$m\\
        $(\sigma_{\mathrm{sg}})^{2}$ & $7.96 \times 10^{-12}$W~\cite{sg} & $\lambda_{\mathrm{g}}$ & $8000$~\cite{pa2}\\
        $(\sigma_{\mathrm{s}})^{2}$ & $10^{-13}$W~\cite{ss} & $d_{\mathrm{sg}}$ & $550$km~\cite{ss}\\
        $f_{\mathrm{sg}}$ & $2$GHz~\cite{sg} & $G_{\mathrm{sg}}$ & $38$dBi~\cite{sg}\\
        $b_{0}$ & $0.851$~\cite{sg} & $m_{0}$ & $2.91$~\cite{sg}\\
        $B_k$ & $\{1\mathrm{kHz},1\mathrm{kHz},20\mathrm{MHz}\}$& $\Omega$ & $0.278$~\cite{sg}\\
         $C_k$ & $\{\mathrm{8kbps,8kbps,200Mbps}\}$ & $\eta_{s}$ & $1.00526$\cite{ss}\\
        $R_k$ & $\{\mathrm{5kbps,3kbps,80Mbps}\}$ & $G_{\mathrm{S}}$ & $160$dBi~\cite{ss} \\
        $A_{0}$ & $0.01979$~\cite{ss} & $\varsigma$ & $15$mrad~\cite{ss}\\
        $\gamma_{0,k}$ & $-100$dBm & $\varepsilon$ & $10^{-14}$\\
        
		\hline
	\end{tabular}
\end{table}

\section{Numerical Results}
\label{sec_results}
\begin{figure}[t]
	\captionsetup[subfigure]{justification=centering}
	\centering
	\subfloat[\\$n=5$
    ]{\includegraphics[width=4.3cm]{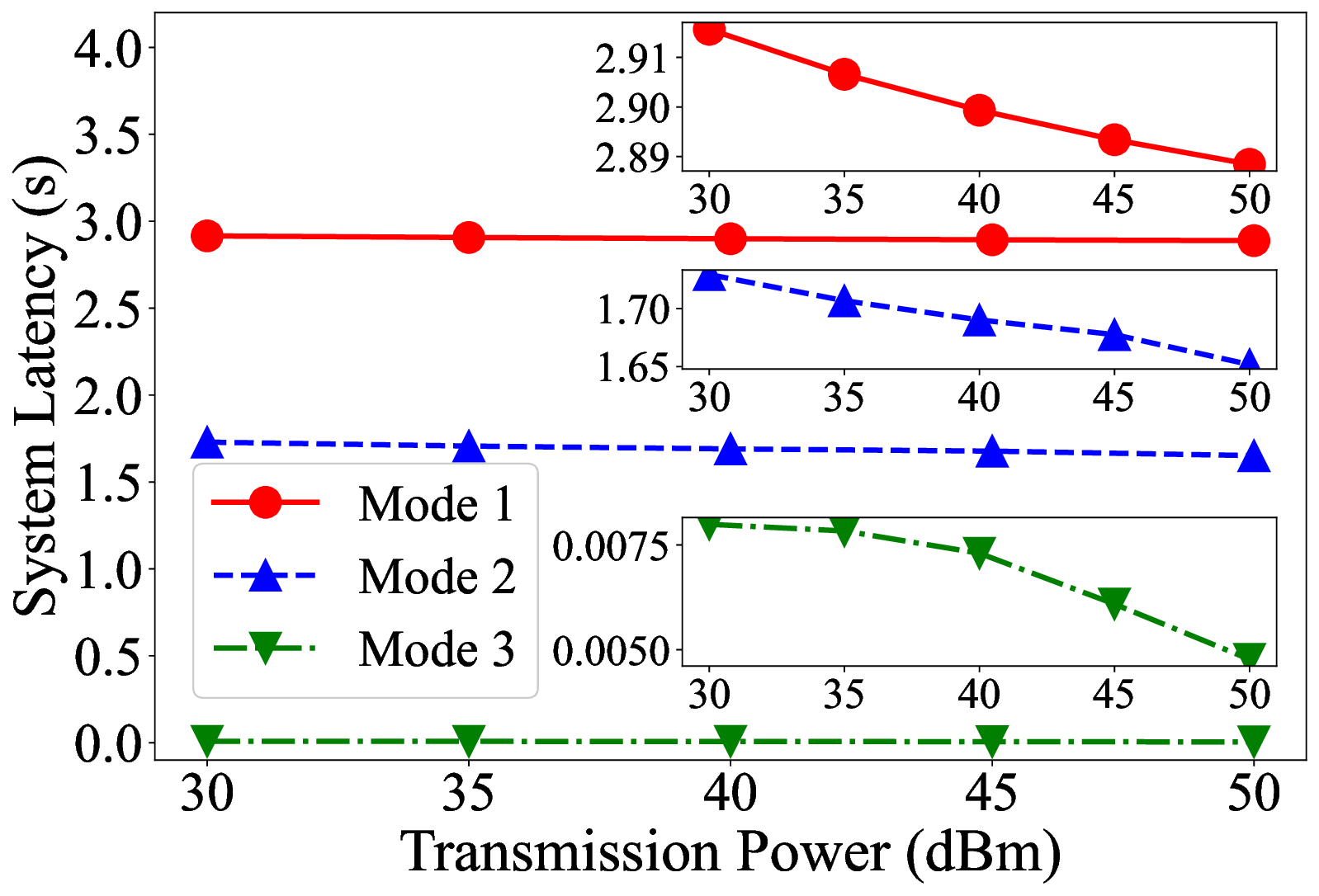}\label{fig3_1}}	
     \hfil 
    \subfloat[ $\{p_{\mathrm{g}},p_{\mathrm{sg}},p_{\mathrm{s}}\}=\{40,40,40\}$dBm
    ]{\includegraphics[width=4.15cm]{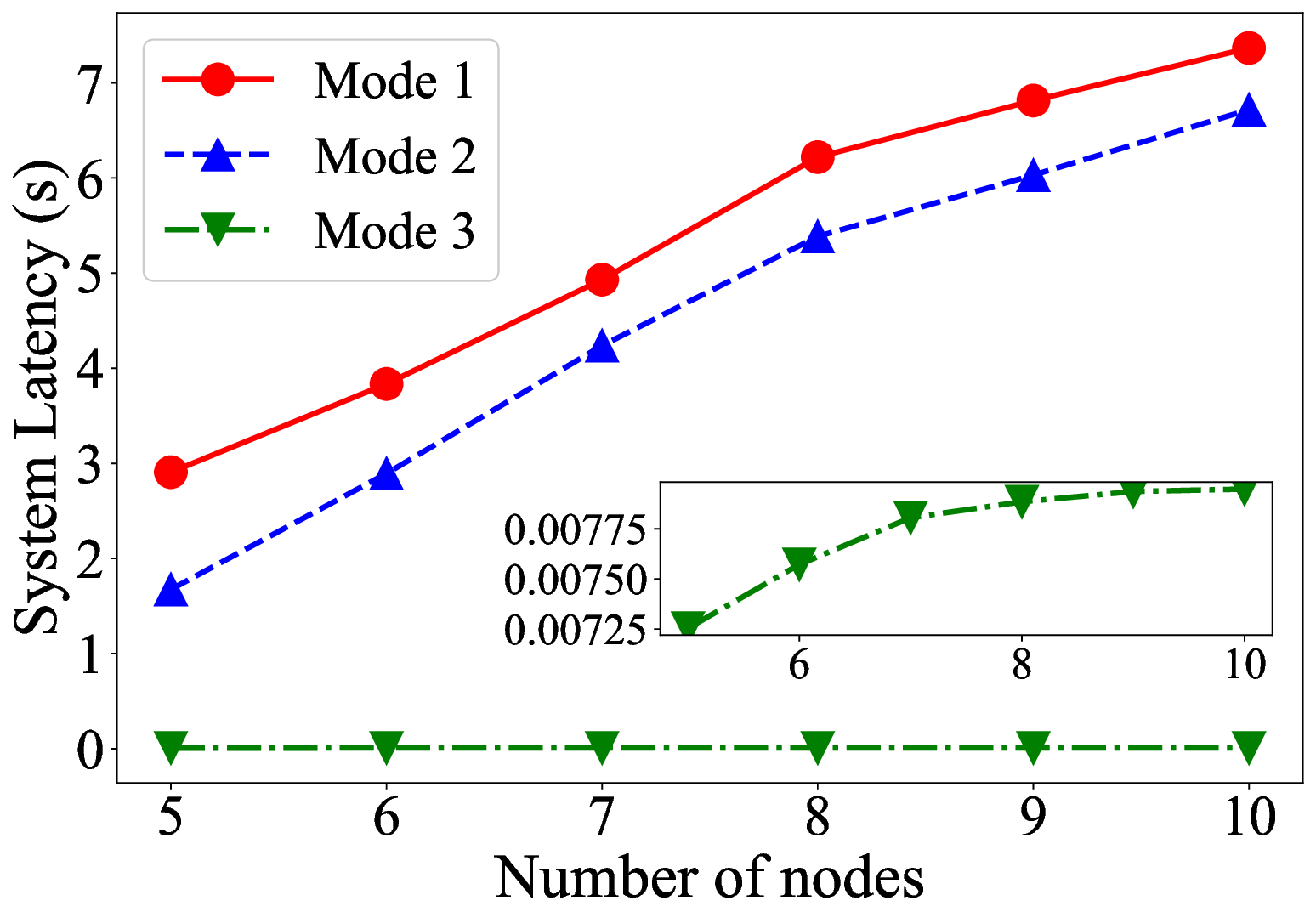}\label{fig3_2}}	\hfil
	\caption{Simulation results of system latency versus transmission power $\{p_{\mathrm{g}},p_{\mathrm{sg}},p_{\mathrm{s}}\}$ and number of chain nodes $n$ in three hybrid modes.}
	\label{fig_simu}
\end{figure}

In this section, we conduct numerical experiments to analyze the system performance for the proposed DID chain in three hybrid modes. Considering $n$ number of chain nodes for each mode, in \textbf{Mode 1} and \textbf{Mode 3}, all $n$ chain nodes are ground nodes or satellite nodes, and in \textbf{Mode 2}, $n-1$ ground nodes and $1$ satellite node are used. To verify the effectiveness of our analytical value ranges of $T_{\mathrm{overall}}$ and $\mathsf{TPS}$ in~\cref{subsec_latencyanalysis}, we give both simulation and theoretical results under different system parameters. Since the throughput $\mathsf{TPS}$ is the reciprocal of the system latency $T_{\mathrm{overall}}$, we just show the results of $T_{\mathrm{overall}}$ and its analysis can also reflect the performance on $\mathsf{TPS}$. Each simulated system latency is generated by averaging $1000$ Monte Carlo simulations, and each simulation uses random node distributions, propagations, and pointing errors catering to the real-world environments. The simulation parameters are given in Table I. Since the credentials size of DID system is often less than 1kb \cite{did2}, the size transmission bits is set as 128 bits in our simulations. Besides, the values of $\{B_{k},C_{k},R_{k}\}$ refer to \cite{ps1} and \cite{isl}. \cref{fig_simu} gives simulation results of system latency under three hybrid modes. \cref{fig_simu}\subref{fig3_1} and \cref{fig_simu}\subref{fig3_2} investigate the impact of the changing transmission powers and chain nodes in each mode on system latency, respectively. In \cref{fig_simu}\subref{fig3_1}, the system latency of all modes decreases with the increase of the transmission power. This is because a larger transmission power leads to a higher SNR/SINR, which in turn increases the transmission rates. In \cref{fig_simu}\subref{fig3_2}, the system latency of all modes increases with the increase of the total number of chain nodes. This is because there are more links in a system with more nodes, which may slow down the time it takes for PBFT to reach the consensus. Comparing the results of all three modes, we find that \textbf{Mode 1} always has the longest latency, while \textbf{Mode 3} always completes the fastest. The reason is that LEO satellites use high-speed ISLs to communicate, which are much faster than traditional satellite-to-ground links because of the larger bandwidth; meanwhile, the communication in \textbf{Mode 1} faces interference from other ground nodes.

\begin{figure*}[t]
	\centering
    \vspace{-1mm}
	\subfloat[Mode 1, $n=6,9$
    ]{\includegraphics[width=5.1cm]{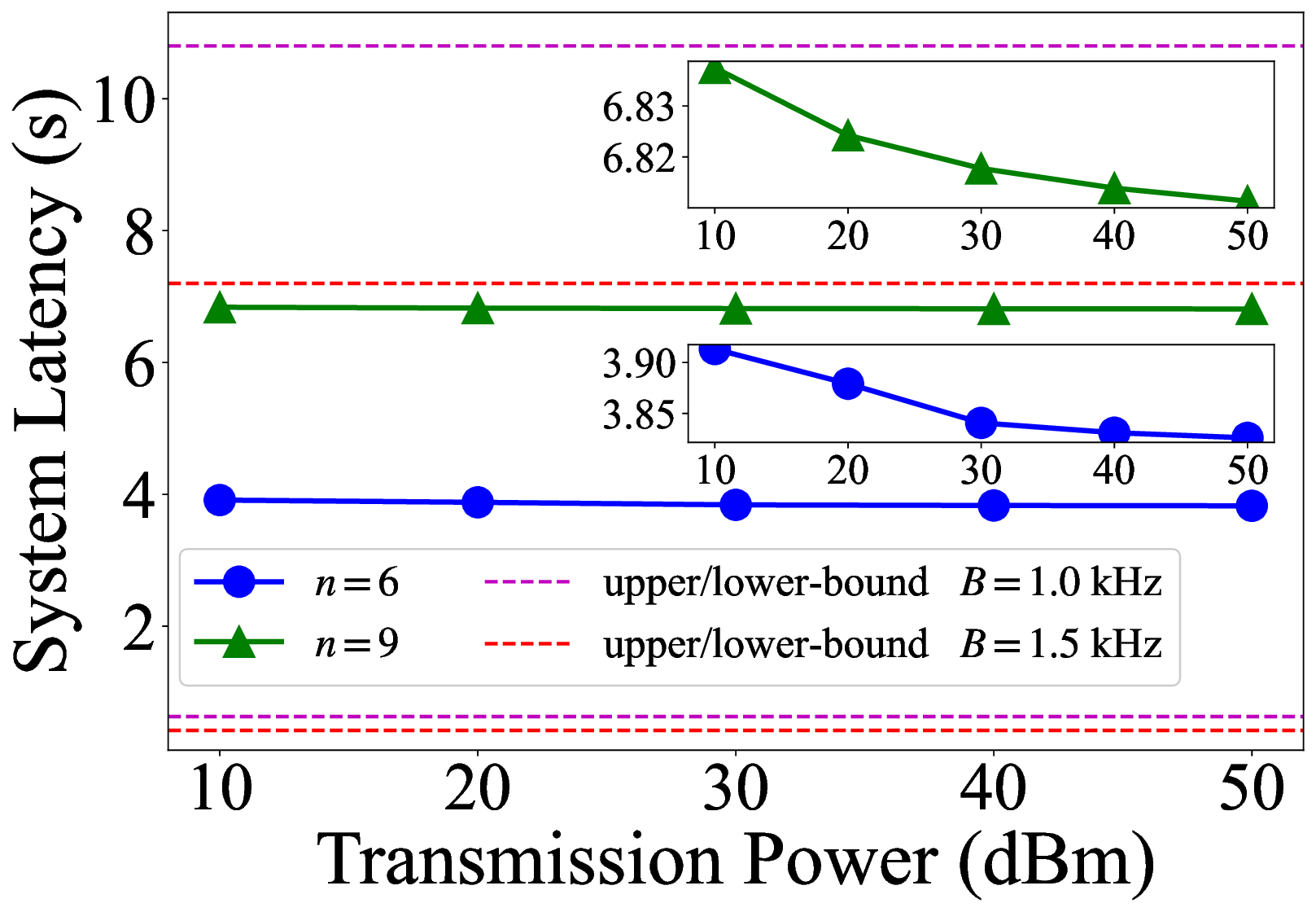}\label{fig4_1}}	\hfil
	\subfloat[Mode 2, $n=6,9$
    ]{\includegraphics[width=5.15cm]{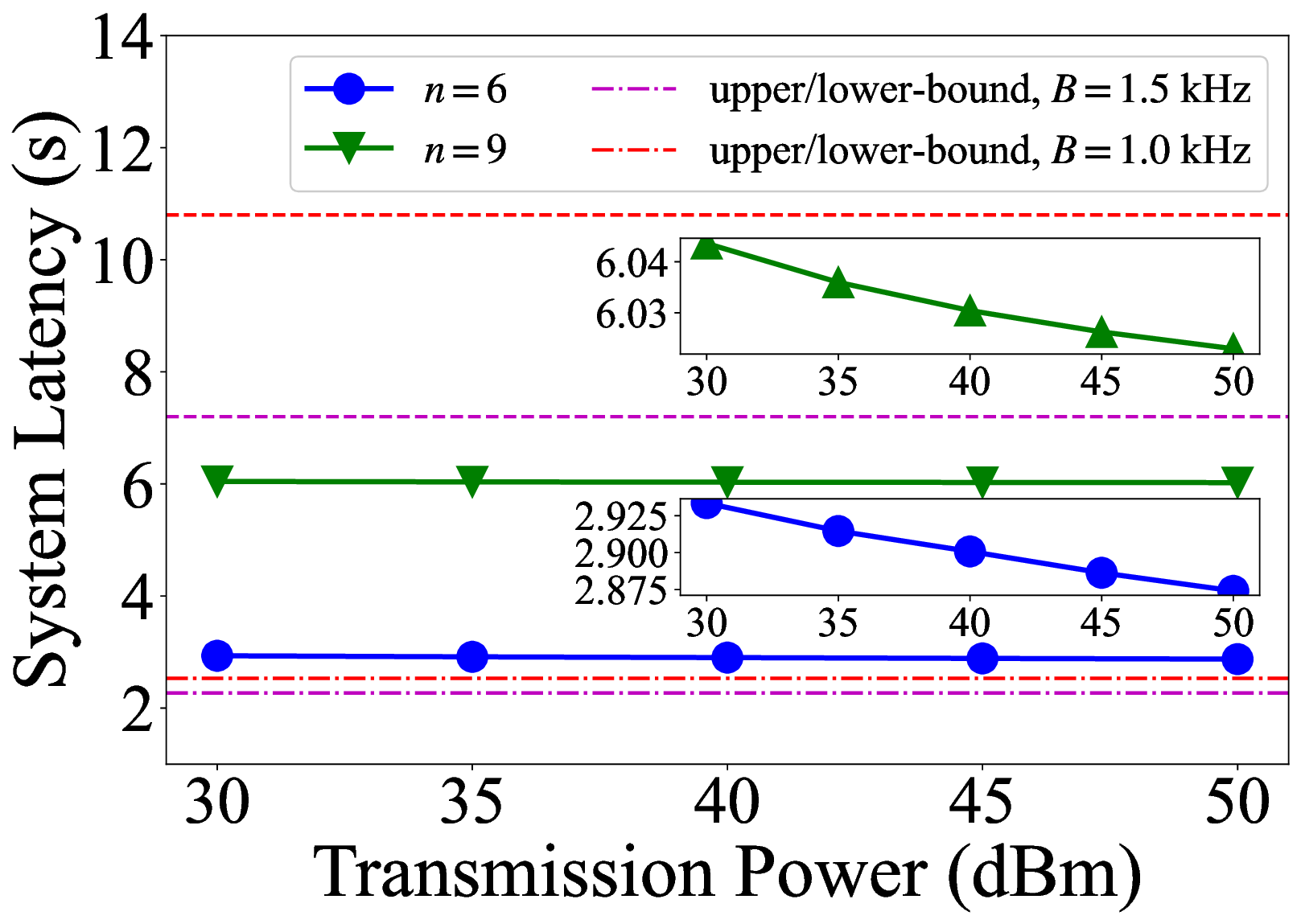}\label{fig4_2}}	\hfil
	\subfloat[Mode 3, $n=6,9$
    ]{\includegraphics[width=5.05cm]{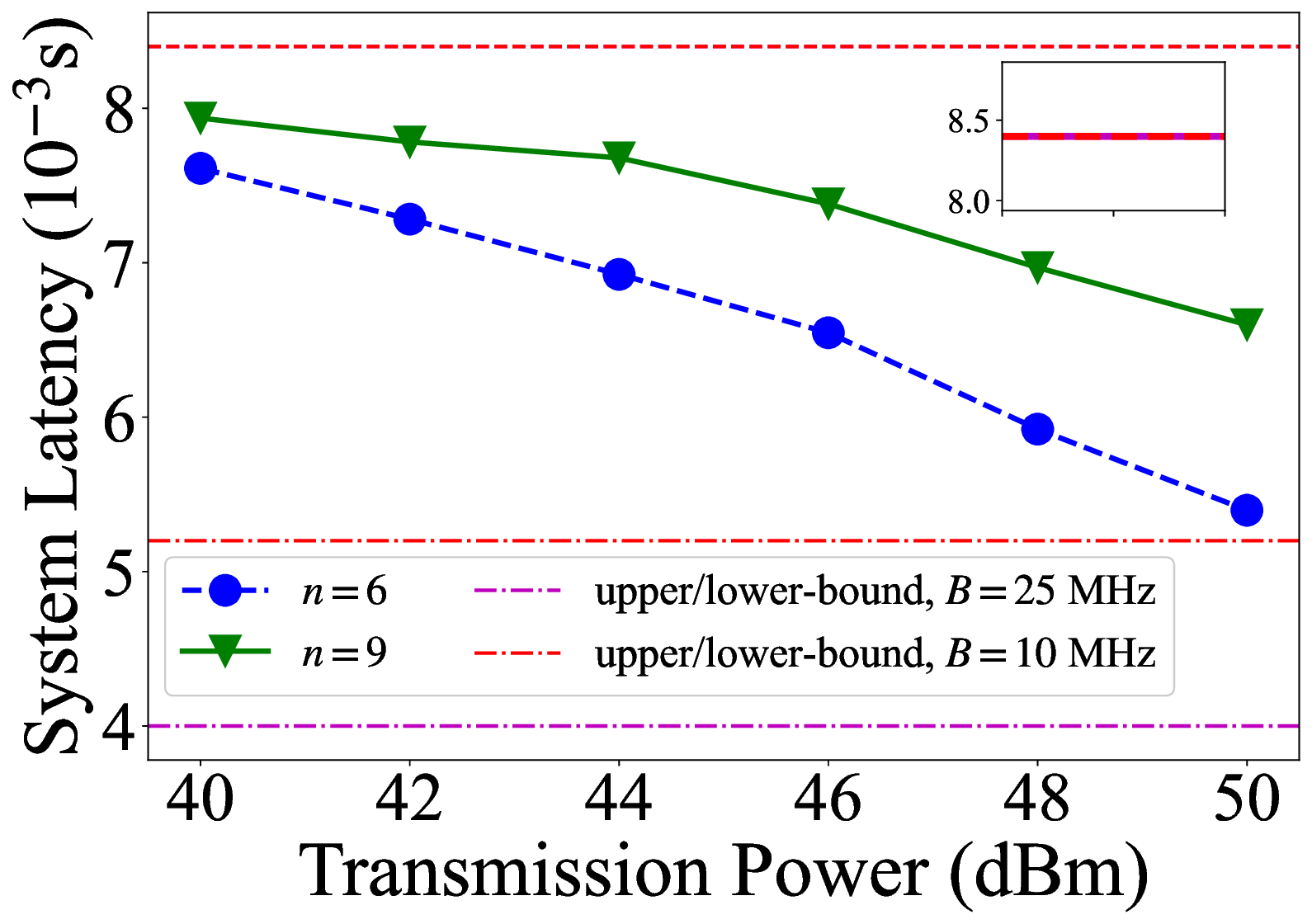}\label{fig4_3}}	\hfil
	\caption{System latency versus transmission power $\{p_{\mathrm{g}},p_{\mathrm{sg}},p_{\mathrm{s}}\}$.}
	\label{fig_theo}
\end{figure*}

\cref{fig_theo} shows both simulation results and the theoretical value ranges of system latency for three modes. Particularly, we depict the changing system latency with the different transmission power of chain nodes and different numbers of nodes ($n=6,9$). From~\cref{fig_theo}, we see that all the simulation results of the latency of the three modes satisfy the value range of the theoretical analysis; this verifies the effectiveness of our theoretical modeling. Comparing~\cref{fig_theo}\subref{fig4_1} and~\cref{fig_theo}\subref{fig4_2}, the upper bound of the system latency in \textbf{Mode 1} and \textbf{Mode 2} are the same and determined by the inter-ground links since communications between ground nodes are often subject to interference. Moreover, compared with \textbf{Mode 1}, the latency of \textbf{Mode 2} has a larger lower bound since the satellite-ground links suffer a more serious channel fading and path loss.

\cref{fig_theo} also presents the upper/lower bounds of the latency of three modes with different bandwidths. This means that the system takes less time to achieve consensus when the transmission links have a larger bandwidth. The trends of these derived bounds are consistent with the results of the theoretical analysis in Theorem 1. Moreover, \cref{fig_theo} shows that system latency decreases with increasing transmission power when $n=6$ and $9$. 
Specifically, ~\cref{fig_theo}\subref{fig4_3} indicates that, in \textbf{Mode 3}, the impact of transmission power variation on system latency is more obvious when the number of nodes is small because the system has fewer redundant paths for communication, when the number of nodes is small, which makes transmission power fluctuations more critical to performance. On the contrary, although the system latency of \textbf{Mode 1} and \textbf{Mode 2} shown in~\cref{fig_theo}\subref{fig4_1} and~\cref{fig_theo}\subref{fig4_2} also show the same trend, their latency variations are not very different in terms of sensitivity to transmission power under different $n$.

\noindent 
\textbf{Insights:} The results provide some practical insights for deploying Web3 DID in hybrid satellite-terrestrial networks.
\vspace{-0.2cm}

{\footnotesize
\begin{itemize}[leftmargin=*]
    \item The simulated system latency of all modes decreases with the increase of the transmission power and the decrease of the total number of chain nodes. This shows the effectiveness of the proposed deployment modes under different real-world environments. \textbf{Mode 1} always has the longest latency because of crowded inter-ground links, while \textbf{Mode 3} always completes the fastest because of high speed and stable ISL links. 
    \item All the simulation results satisfy the value range of the theoretical analysis; this verifies the effectiveness of our theoretical results. 
    Compared with \textbf{Mode 1}, the latency of \textbf{Mode 2} has a larger lower bound since the satellite-ground links suffer a more serious channel fading and path loss. In \textbf{Mode 3}, the impact of transmission power variation on system latency is more obvious when the number of nodes is small because the system has fewer redundant paths for communication, which makes transmission power fluctuations more critical to performance. 
\end{itemize}
}

\section{Conclusions}
\label{sec_conclusion}
\vspace{-0.3em}
We propose hybrid satellite-ground deployment modes to enhance the coverage and reliability of Web3 DID systems. Through a theoretical performance analysis, we derive trackable expressions for the upper and lower bounds of latency for three different modes, providing insights into the impact of wireless network factors on blockchain consensus and DID management. Our simulation results validate the effectiveness of the proposed deployment modes under real-world environments. The observations of results give practical insights for deploying Web3 DID in hybrid satellite-terrestrial networks.
\vspace{-0.3em}

{\appendices
\section{}
\label{proof_lem1}
\vspace{-0.3em}
\textit{Proof of Lemma 1:}
In
~\cref{subsubsec_DIDchain}, the first three phases of PBFT consensus require nodes to broadcast messages to the other nodes. Then the phase latency $T_i(i=\{1,2,3\})$ depends on the maximum latency of the communication between any two nodes, i.e., $t_c(x_{\mathrm{t}},x_{\mathrm{r}})$. In the last phase of PBFT consensus, the consensus is completed after all nodes send the \texttt{REPLY} message back. Thus the phase latency $t_4$ is the latency of the communication from any node $x_{\mathrm{t}}\in\mathcal{X}$ and the primary node $x_{\mathrm{0}}$. Thereby, we have
~\cref{lem1}.\hfill $\blacksquare$

\vspace{-0.3em}
\section{}
\label{proof_lem2}
\vspace{-0.3em}
\textit{Proof of Lemma 2:} Given a wireless communication system including the parameters $\{N,B,C,R\}$, the relationship between $t_c$ and $P_{s}$ follows the formula below~\cite{pa1,ps1,ps2}.
\vspace{-0.3cm}

{\footnotesize
\begin{equation*}
    1-P_{s}=Q\left(\frac{(C-R)BNt_c+0.5\log_{2}(BNt_c)}{\log_{2}e\cdot \sqrt{BNt_c}}\right),
\end{equation*}}%
where $Q(\cdot)$ is the $Q$-function.
Let the independent variable part of $Q(\cdot)$ in the above equation as a new function $\Phi(t_c)$ with respect to $t_c$. 
Furthermore, from Equaition (4) in \cite{ps2}, $\Phi(t_c)$ is a monotonic increasing function of $t_c$, it must have a unique corresponding inverse function $\Phi^{-1}(t_c)$. Accordingly, $t_c$ can be calculated in~\cref{lem2}.\hfill$\blacksquare$ 
\vspace{-0.1cm}
\section{}
\label{proof_pro1}
\vspace{-0.3em}
\textit{Proof of Proposition 1:} The successful transmission probability $P_{s,l}$ of a link is usually determined by the probability of the received SNR/SINR $\gamma_{l}$ over a minimum threshold $\gamma_{0,l}$ that the receiver can decode the signal, thus, $\mathbb{P}(\gamma_{l}\geq \gamma_{0,l})$. Among the three types of transmission links, some are robust to interference and some are not. Therefore, the calculation of $P_{s}$ includes two cases, i.e., interference-free and interference-involved. For the interference-free case, $\gamma_{l}$ is SNR, and no interference is involved. For the interference case, $\gamma_{l}$ is SINR, and the probability calculation needs to consider the distribution of interferers, i.e., related to the distance distribution $d$ from all other ground nodes to the receiver. To sum up, we have the general formula of $P_{s,l}$ in~\cref{pro0}.
\hfill $\blacksquare$

\vspace{-0.1cm}
\section{}
\label{proof_cro1}
\vspace{-0.3em}
\textit{Proof of Corollary 1:} 
Substituting~\cref{eq_gamma_g,eq_fi} into~\cref{pro0}, we have
\vspace{-0.5cm}

{\footnotesize
\begin{align}
    &P_{s,\mathrm{g}}=
    \mathbb{P}(\gamma_{\mathrm{g}}\geq \gamma_{0,\mathrm{g}})
    \overset{(a)}{=}\mathbb{P}\left(\int_{\gamma_{0,\mathrm{g}}-\varepsilon}^{+\infty}f_{\gamma_{\mathrm{g}}}\mathrm{d}\gamma_{\mathrm{g}}\geq \int_{\gamma_{0,\mathrm{g}}-\varepsilon}^{+\infty}\gamma_{0,\mathrm{g}}'\mathrm{d}\gamma_{\mathrm{g}}\right)\notag \\
    &=\mathbb{P}\left(\prod\limits_{k=1}^{K}d_x\geq \frac{\gamma_{0,\mathrm{g}}'}{f_{S_{\mathrm{g}}}}\left(\frac{D_{\mathrm{g}}^{2}}{2}\right)^{K}\right),
    \label{a1}
\end{align}}%
where $\varepsilon$ is a constant and satisfies $0<\varepsilon<\gamma_{0,\mathrm{g}}$ and $(a)$ comes from the substitution of
\vspace{-0.3cm}

{\footnotesize
\begin{align}
    &\gamma_{0,\mathrm{g}}'=\frac{1-\int_{\gamma_{0,\mathrm{g}}-\varepsilon}^{\gamma_{0,\mathrm{g}}}f_{\gamma_{\mathrm{g}}}\mathrm{d}\gamma_{\mathrm{g}}}{\gamma_{0,\mathrm{g}}-\varepsilon},
    \int_{\gamma_{0,\mathrm{g}}-\varepsilon}^{\gamma_{0,\mathrm{g}}}f_{\gamma_{\mathrm{g}}}\mathrm{d}\gamma_{\mathrm{g}}=\int_{\gamma_{0,\mathrm{g}}-\varepsilon}^{+\infty}\gamma_{0,\mathrm{g}}'\mathrm{d}\gamma_{\mathrm{g}}.\label{aa2}
\end{align}}
In~\cref{a1}, each $d_x$ in $\prod_{k=1}^{K}d_x$ is a distance random variable from an interferer node to the desired receiver. Let $d_{\min}$ and $d_{\max}$ be the minimum and maximum value of $d_x$. Due to the homogeneity of ground nodes and the value range of a Cumulative Density Function (CDF) of the random variable $\prod_{k=1}^{K}d_x$,
we have the upper bound $P_{s,\mathrm{g}}^{\mathrm{L}}=\mathbb{P}\left(d_{\min}^{K}\geq \gamma_{0,\mathrm{g}}'/f_{S_{\mathrm{g}}}\cdot\left(D_{\mathrm{g}}^{2}/2\right)^{K}\right)$ and lower bound $P_{s,\mathrm{g}}^{\mathrm{U}}=\mathbb{P}\left(d_{\max}^{K}\geq \gamma_{0,\mathrm{g}}'/f_{S_{\mathrm{g}}}\cdot\left(D_{\mathrm{g}}^{2}/2\right)^{K}\right)$. Let $\gamma_{0,L}'$ and $\gamma_{0,U}'$ be the lower bound and upper bound of $\gamma_{0,\mathrm{g}}$. Referring to~\cref{aa2}, we have $ \gamma_{0,L}'\leq\gamma_{0,\mathrm{g}}'\leq\gamma_{0,U}',\gamma_{0,L}'=(1-\varepsilon)/(\gamma_{0,\mathrm{g}}-\varepsilon),\gamma_{0,U}'=1/(\gamma_{0,\mathrm{g}}-\varepsilon)$. Combining $\gamma_{0,L}',\gamma_{0,U}',P_{s,\mathrm{g}}^{\mathrm{L}},P_{s,\mathrm{g}}^{\mathrm{U}}$, we have
\vspace{-0.3cm}

{\footnotesize
\begin{align*}
    &P_{s,\mathrm{g}}^{\mathrm{L}}
    \geq \mathbb{P}\left(d_{\min}^{K}\geq \frac{\gamma_{0,U}'}{f_{S_{\mathrm{g}}}}\left(\frac{D_{\mathrm{g}}^{2}}{2}\right)^{K}\right)
    =1-\mathbb{P}\left(d_{\min}\leq \left(\frac{\gamma_{0,U}'}{f_{S_{\mathrm{g}}}}\right)^{\frac{1}{K}}\frac{D_{\mathrm{g}}^{2}}{2}\right)\notag\\
    &\overset{(a)}{=}1-\left(1-\left(1-\frac{1}{D_{l}^{2}}\left(\frac{\gamma_{0,U}'}{f_{S_{\mathrm{g}}}}\right)^{\frac{2}{K}}\frac{D_{\mathrm{g}}^{4}}{4}\right)^{K}\right)
    =\left(1-\frac{D_{\mathrm{g}}^{2}}{4}\left(\frac{\gamma_{0,U}'}{f_{S_{\mathrm{g}}}}\right)^{\frac{2}{K}}\right)^{K},\\&P_{s,\mathrm{g}}^{\mathrm{U}}
    \leq \mathbb{P}\left(d_{\max}^{K}\geq \frac{\gamma_{0,L}'}{f_{S_{\mathrm{g}}}}\left(\frac{D_{\mathrm{g}}^{2}}{2}\right)^{K}\right)
    =1-\mathbb{P}\left(d_{\max}\leq \left(\frac{\gamma_{0,L}'}{f_{S_{\mathrm{g}}}}\right)^{\frac{1}{K}}\frac{D_{\mathrm{g}}^{2}}{2}\right)\notag\\
    &\overset{(b)}{=}1-\left(\frac{1}{D_{\mathrm{g}}^{2}}\left(\frac{\gamma_{0,L}'}{f_{S_{\mathrm{g}}}}\right)^{\frac{2}{K}}\frac{D_{\mathrm{g}}^{4}}{4}\right)^{K}
    =1-\frac{D_{\mathrm{g}}^{2K}}{4^{K}}\left(\frac{\gamma_{0,L}'}{f_{S_{\mathrm{g}}}}\right)^{2},
\end{align*}
}
where $(a)$ and $(b)$ comes from $\mathbb{P}(X\leq x)=F_{X}(X)$. 
Accordingly, we have the new lower bound and upper bound of $P_{s,\mathrm{g}}$. 
Substituting $f_{S_{\mathrm{g}}}=2\pi\lambda_{\mathrm{g}}d_0e^{-\lambda_{\mathrm{g}}\pi d_0^{2}}$ and $K=\left \lceil \pi \lambda_{\mathrm{g}} D_{\mathrm{g}}^2\right \rceil$ into bound functions, we have~\cref{cro1}.
\hfill$\blacksquare$
\vspace{-0.3em}
\section{}
\label{proof_cro2}
\vspace{-0.3em}
\textit{Proof of Corollary 2:} Substituting~\cref{eq_gamma_sg,eq_fh} into~\cref{pro0}, $P_{s,\mathrm{sg}}$ can be calculated by $P_{s,\mathrm{sg}}=1-\mathbb{P}\left(|h_{\mathrm{sg}}|^{2}\leq \frac{(\sigma_{\mathrm{sg}})^{2}\gamma_{0,\mathrm{sg}}}{p_{\mathrm{sg}}G_{\mathrm{sg}}L_{\mathrm{sg}}(d_0)}\right)$. 
Substituting the CDF of $|h_{\mathrm{sg}}|^{2}$ into the above equation, we have~\cref{cro2}.
\hfill$\blacksquare$
\section{}
\label{proof_cro3}
\textit{Proof of Corollary 3:}
Substituting~\cref{eq_gamma_s,eq_fw} into~\cref{pro0}, $P_{s,\mathrm{s}}$ can be calculated as follow. 

{\footnotesize
\begin{align}&P_{s,\mathrm{s}}=\mathbb{P}\left(\gamma_{\mathrm{s}}\geq\gamma_{0,\mathrm{s}}\right)\overset{(a)}{=}(1-\varsigma^{2})\left(1-\left(\frac{\gamma_{0,\mathrm{s}}\sigma_{\mathrm{s}}^{2}16\pi^2 c^2d_0^{2}}{A_{0}p_{\mathrm{s}}G_{\mathrm{s}}f_{\mathrm{s}}^{2}}\right)^{\eta_{s}^{2}}\right),\notag
\end{align}}%
where $(a)$ follows the similar derivation of the Lemma 1 in \cite{ss}. Given any two satellites on the altitudes $h_{i}$ and $h_{j}$, the maximum visible distance is $d_{0}^{\max}=\sqrt{h_{i}(h_{i}+2R_{\mathrm{e}})}+\sqrt{h_{j}(h_{j}+2R_{\mathrm{e}})}$ with $R_{\mathrm{e}}$ being the earth radius. Then 
$P_{s,\mathrm{s}}$ has the lower bound and upper bound in~\cref{cro3}.
\hfill$\blacksquare$
\section{}
\label{proof_the1}
\textit{Proof of Theorem 1:} In \textbf{Mode 1} ($k=1$), the DID consensus is completed via \textit{inter-ground links} among ground chain nodes. Referring to~\cref{lem2}, given $P_{s,\mathrm{g}}$ of a inter-ground link, the latency $t_{c,\mathrm{g}}$ can be calculated by $t_{c,\mathrm{g}}=\Phi^{-1}\left(Q^{-1}(1{-}P_{s,\mathrm{g}})\right)$
In this equation, $Q^{-1}(\cdot)$ is a monotonic decreasing function, $\Phi^{-1}(\cdot)$ is 
a monotonic increasing function, then $t_{c,\mathrm{g}}$ have monotonic trend relationship with $P_{s,\mathrm{g}}$. Substituting~\cref{cro1} into~\cref{lem2}, we have the lower bound and upper bound of $t_{c,\mathrm{g}},T_{1,\mathrm{overall}}$ as follows.
\vspace{-0.3cm}

{\footnotesize
\begin{align*}
        &4t_{c,\mathrm{g}}^{\mathrm{L}}\leq T_{1,\mathrm{overall}}\leq 4t_{c,\mathrm{g}}^{\mathrm{U}},t_{c,\mathrm{g}}^{\mathrm{L}}\leq t_{c,\mathrm{g}} \leq t_{c,\mathrm{g}}^{\mathrm{U}},\\
        &t_{c,\mathrm{g}}^{\mathrm{L}}=\Phi^{-1}\left(Q^{-1}(1{-}P_{s,\mathrm{g}}^{\mathrm{U}})\right),
        t_{c,\mathrm{g}}^{\mathrm{U}}=\Phi^{-1}\left(Q^{-1}(1{-}P_{s,\mathrm{g}}^{\mathrm{L}})\right).
\end{align*}}%
In \textbf{Mode 2} ($k=2$), the DID consensus is completed via \textit{inter-ground links} among ground chain nodes and \textit{satellite-ground links} between ground chain nodes and satellites. In \textbf{Mode 3} ($k=3$), the DID consensus is completed via \textit{inter-satellite links} among satellite chain nodes. Following the derivation process of $T_{1,\mathrm{overall}}$, we can calculate the lower bound and upper bound of $T_{2,\mathrm{overall}}$ and $T_{3,\mathrm{overall}}$ as follows.
\vspace{-0.3cm}

{\footnotesize
\begin{align*}
        &4\max\{t_{c,\mathrm{g}}^{\mathrm{L}},t_{c,\mathrm{sg}}\}\leq T_{2,\mathrm{overall}}\leq 4\max\{t_{c,\mathrm{g}}^{\mathrm{U}},t_{c,\mathrm{sg}}\},
        \\
        &t_{c,\mathrm{sg}}=\Phi^{-1}\left(Q^{-1}(1{-}P_{s,\mathrm{sg}})\right);\\
        &4t_{c,\mathrm{s}}^{\mathrm{L}}\leq T_{3,\mathrm{overall}}\leq 4t_{c,\mathrm{s}}^{\mathrm{U}},\\
        &t_{c,\mathrm{s}}^{\mathrm{L}}=\Phi^{-1}\left(Q^{-1}(1{-}P_{s,\mathrm{s}}^{\mathrm{U}})\right),
        t_{c,\mathrm{s}}^{\mathrm{U}}=\Phi^{-1}\left(Q^{-1}(1{-}P_{s,\mathrm{s}}^{\mathrm{L}})\right).
\end{align*}}%
The values of $P_{s,\mathrm{sg}},P_{s,\mathrm{s}}^{\mathrm{L}},P_{s,\mathrm{s}}^{\mathrm{U}}$ are given in~\cref{cro2,cro3}. Substituting the value range of $T_{k,\mathrm{overall}}$ into~\cref{lem1}, we can derive the lower bound and upper bound of $\mathsf{TPS}_{k}$ of three modes. Thereby, we have~\cref{the1}.\hfill$\blacksquare$
}

\vspace{-0.4em}
\bibliographystyle{ieeetr}
\bibliography{ref_main}

\begin{thebibliography}{10}

\bibitem{web3}
W.~Liu, B.~Cao, and M.~Peng, ``Web3 {T}echnologies: Challenges and opportunities,'' {\em IEEE Network}, vol.~38, no.~3, pp.~187--193, 2024.

\bibitem{defi}
N.~Naik and P.~Jenkins, ``Sovrin network for decentralized digital identity: Analysing a self-sovereign identity system based on distributed ledger technology,'' in {\em Proc. IEEE International Symposium on Systems Engineering (ISSE)}, pp.~1--7, 2021.

\bibitem{web3b}
J.~Zhu, F.~Li, and J.~Chen, ``A survey of blockchain, artificial intelligence, and edge computing for {W}eb 3.0,'' {\em Computer Science Review}, vol.~54, p.~100667, 2024.

\bibitem{did1}
H.~Deng, J.~Liang, C.~Zhang, X.~Liu, L.~Zhu, and S.~Guo, ``Future{DID}: {A} fully decentralized identity system with multi-party verification,'' {\em IEEE Trans. Comput.}, vol.~73, no.~8, pp.~2051--2065, 2024.

\bibitem{did2}
J.~Yin, Y.~Xiao, Q.~Pei, Y.~Ju, L.~Liu, M.~Xiao, and C.~Wu, ``Smart{DID}: {A} novel privacy-preserving identity based on blockchain for {IoT},'' {\em IEEE Internet Things J.}, vol.~10, no.~8, pp.~6718--6732, 2023.

\bibitem{did3}
Y.~Liu, B.~Zhao, Z.~Zhao, J.~Liu, X.~Lin, Q.~Wu, and W.~Susilo, ``{SS-DID}: {A} secure and scalable web3 decentralized identity utilizing multilayer sharding blockchain,'' {\em IEEE Internet Things J.}, vol.~11, no.~15, pp.~25694--25705, 2024.

\bibitem{s2}
N.~Mohan, A.~E. Ferguson, H.~Cech, R.~Bose, P.~R. Renatin, M.~K. Marina, and J.~Ott, ``A multifaceted look at {S}tarlink performance,'' in {\em Proc. of the ACM Web Conference 2024}, WWW '24, (New York, NY, USA), p.~2723–2734, Association for Computing Machinery, 2024.

\bibitem{pow}
A.~Gervais, G.~O. Karame, K.~W\"{u}st, V.~Glykantzis, H.~Ritzdorf, and S.~Capkun, ``On the security and performance of proof of work blockchains,'' in {\em Proc. of ACM SIGSAC Conference on Computer and Communications Security}, CCS '16, (New York, NY, USA), p.~3–16, Association for Computing Machinery, 2016.

\bibitem{pos}
J.~Spasovski and P.~Eklund, ``Proof of stake blockchain: Performance and scalability for groupware communications,'' in {\em Proc. of the 9th International Conference on Management of Digital EcoSystems}, MEDES '17, (New York, NY, USA), p.~251–258, Association for Computing Machinery, 2017.

\bibitem{pbft1}
Y.~Meshcheryakov, A.~Melman, O.~Evsutin, V.~Morozov, and Y.~Koucheryavy, ``On performance of {PBFT} blockchain consensus algorithm for iot-applications with constrained devices,'' {\em IEEE Access}, vol.~9, pp.~80559--80570, 2021.

\bibitem{pa1}
H.~Luo, X.~Yang, H.~Yu, G.~Sun, B.~Lei, and M.~Guizani, ``Performance analysis and comparison of nonideal wireless {PBFT} and {RAFT} consensus networks in 6{G} communications,'' {\em IEEE Internet Things J.}, vol.~11, no.~6, pp.~9752--9765, 2024.

\bibitem{pa2}
Y.~Sun, L.~Zhang, G.~Feng, B.~Yang, B.~Cao, and M.~A. Imran, ``Blockchain-enabled wireless {I}nternet of {T}hings: Performance analysis and optimal communication node deployment,'' {\em IEEE Internet Things J.}, vol.~6, no.~3, pp.~5791--5802, 2019.

\bibitem{pa3}
F.~Guo, F.~R. Yu, H.~Zhang, H.~Ji, M.~Liu, and V.~C.~M. Leung, ``Adaptive resource allocation in future wireless networks with blockchain and mobile edge computing,'' {\em IEEE Trans. Wireless Commun.}, vol.~19, no.~3, pp.~1689--1703, 2020.

\bibitem{s3}
H.~Wei, W.~Feng, C.~Zhang, Y.~Chen, Y.~Fang, and N.~Ge, ``Creating efficient blockchains for the {I}nternet of {T}hings by coordinated satellite-terrestrial networks,'' {\em IEEE Wireless Commun.}, vol.~27, no.~3, pp.~104--110, 2020.

\bibitem{sg}
Z.~Wang, M.~Cao, H.~Jiang, B.~Cao, S.~Wang, C.~Sun, and M.~Peng, ``Blockchain-enabled dynamic spectrum sharing for satellite and terrestrial communication networks,'' 2024.

\bibitem{rice}
O.~Y. Kolawole, S.~Vuppala, M.~Sellathurai, and T.~Ratnarajah, ``On the performance of cognitive satellite-terrestrial networks,'' {\em IEEE Trans. Cognit. Commun. Networking}, vol.~3, no.~4, pp.~668--683, 2017.

\bibitem{point}
Apoorva, S.~Bitragunta, and S.~Nitundil, ``{Best beam selection and PHY switching policy for hybrid FSO/RF inter-satellite communication link},'' {\em IET Commun.}, vol.~14, no.~19, pp.~3350--3362, 2020.

\bibitem{ss}
R.~Wang, M.~A. Kishk, and M.-S. Alouini, ``Ultra reliable low latency routing in {LEO} satellite constellations: {A} stochastic geometry approach,'' {\em IEEE J. Sel. Areas Commun.}, vol.~42, no.~5, pp.~1231--1245, 2024.

\bibitem{ps1}
X.~Yang, H.~Luo, J.~Duan, and H.~Yu, ``Ultra reliable and low latency authentication scheme for internet of vehicles based on blockchain,'' in {\em Proc. IEEE Conference on Computer Communications Workshops (INFOCOM WKSHPS)}, pp.~1--5, 2022.

\bibitem{isl}
I.~Leyva-Mayorga, B.~Soret, and P.~Popovski, ``Inter-plane inter-satellite connectivity in dense {LEO} constellations,'' {\em IEEE Trans. Wireless Commun.}, vol.~20, no.~6, pp.~3430--3443, 2021.

\bibitem{ps2}
D.~Yu, W.~Li, H.~Xu, and L.~Zhang, ``Low reliable and low latency communications for mission critical distributed industrial internet of things,'' {\em IEEE Commun. Lett.}, vol.~25, no.~1, pp.~313--317, 2021.

\end{thebibliography}

\vfill
\end{document}